# A biosensor based on magnetoelastic waves for detection of antibodies in human plasma for COVID-19 serodiagnosis

Wenderson R. F. Silva[1,*], Larissa C. P. Monteiro[2], Renato L. Senra[2], Eduardo N. D. de Araújo[1], Rafael O. R. R. Cunha[1], Tiago A. O. Mendes[2], Joaquim B. S. Mendes[1,†]

[1]Departamento de Física, Universidade Federal de Viçosa, 36570-900 Viçosa, Minas Gerais, Brazil.
[2]Departamento de Bioquímica e Biologia Molecular, Universidade Federal de Viçosa, 36570-900 Viçosa, Minas Gerais, Brazil.

## Abstract

This study proposes a new efficient wireless biosensor based on magnetoelastic waves for antibody detection in human plasma, aiming at the serological diagnosis of COVID-19. The biosensor underwent functionalization with the N antigen - nucleocapsid phosphoprotein of the SARS-CoV-2 virus. Validation analyses by sodium dodecyl-sulfate polyacrylamide gel electrophoresis (SDS-PAGE), Western blotting (WB), atomic force microscopy (AFM), scanning electron microscopy (SEM), energy-dispersive X-ray (EDX) microanalysis and micro-Raman spectroscopy confirmed the selectivity and effective surface functionalization of the biosensor. The research successfully obtained, expressed and purified the recombinant antigen, while plasma samples from COVID-19 positive and negative patients were applied to test the performance of the biosensor. A performance comparison with the enzyme-linked immunosorbent assays (ELISA) method revealed equivalent diagnostic capacity. These results indicate the robustness of the biosensor in reliably differentiating between positive and negative samples, highlighting its potential as an efficient and low-cost tool for the serological diagnosis of COVID-19. In addition to being fast to execute and having the potential for automation in large-scale diagnostic studies, the biosensor fills a significant gap in existing SARS-CoV-2 detection approaches.

Corresponding authors: *wenderson.f@ufv.br, †joaquim.mendes@ufv.br

## 1. Introduction

Magnetoelastic (ME) materials present a strong coupling between their magnetic and elastic properties, making them an excellent platform for sensor development (O'Handley, 2008). These materials are amorphous and soft ferromagnetic that change their magnetoelastic resonance frequency in response to applied stress, such as the increase in mass on the sensor surface, due to the deposition of a selectively fixed target, for example. ME materials are a prominent platform for manufacturing sensors capable of detecting several physical, chemical and biological parameters (Narita et al., 2020; Saiz et al., 2022). The main advantages are the possibility of high sensitivity and wireless detection (Sang et al., 2020; Liu et al., 2019; Wang et al., 2020), chemically and physically resistant (Sagasti et al., 2019; Sagasti et al., 2018), in addition to being versatile and easily technically viable for production and detection, dispensing with expensive chemical reagents, electrical contacts and lithographic processes. Furthermore, remote analysis capability can offer more advantages where a direct probe or electrical contact with the sensor is not a viable alternative.

Severe acute respiratory syndrome, caused by the SARS-CoV-2 virus, began the COVID-2019 disease pandemic in 2019 with several clinical and geopolitical consequences to this day (Binshaya, 2023; Chen et al. 2022). The COVID-19 pandemic boosted the search for sensitive and accurate detection methods for the SARS-CoV-2 virus and biomarkers such as virus-specific antibodies (Abdelhamid et al., 2021; Kevadiya et al., 2021). Among these methods, the polymerase chain reaction (PCR) for active viral infection detection and viral load quantification and the ELISA for both virus antigen and antibody detection induced by the infection as biomarkers of contact with the virus are currently considered the gold detection method. Besides these, other detection methods have been developed, such as electrochemical sensors (Timilsina et al. 2023; Kumar et al. 2022; Raziq et al., 2021; Yakoh et al., 2021; Zhao et al., 2021; Rashed et al., 2021), capacitive sensors (Georgas et al., 2022; Park et al., 2022), biosensors based on field effect transistors (JFET) (Seo et al., 2020; Li et al., 2021; Shao et al., 2021) and magnetostriction sensors (Neyama et al., 2023). Thus, synthetically, a comparison of several serological methods for detecting antibodies produced in response to SARS-CoV-2 infection is shown in Table 1. However, magnetoelastic sensors have not been reported and tested for the detection of COVID-19 antibodies yet. The absence of reported magnetoelastic sensors in the context of SARS-CoV-2 virus detection highlights a significant gap in existing approaches, mainly due to the possibility of automating the technique to infectious diseases serodiagnosis compared to the ELISA method, allowing analysis of a large number of samples, as necessary in epidemiological studies.

**Table 1** - Comparison of different serological methods for the detection of antibodies produced in response to SARS-CoV-2 infection. Where it says NI (not informed), the authors did not provide this information in their papers.

| Method | Detection | Positive samples | Negative samples | Sensitivity | Specificity | Accuracy | Assay time (min) | Reference |
|---|---|---|---|---|---|---|---|---|
| EEVD | Serologic IgG | 9 | 9 | 66.7 % | NI | 61.1 % | 15 | Mattioli et al. (2022) |
| Colorimetric | Serologic IgG | 10 | 10 | 83 % | 100 % | NI | 30 | Lew et al. (2021) |
| Electrochemical | Serologic IgG and IgM | 7 | 10 | 100 % | 90 % | NI | 30 | Yakoh et al. (2021) |
| Electrochemical | Serologic IgG | 54 | 39 | 100 % | 100 % | NI | < 10 | Timilsina et al. (2023) |
| Microfluidic Serological | Serologic IgG and IgM | 60 | 92 | 91.7 % | 100 % | 96.7 % | 5 | Lee et al. (2021) |
| Rapid point-of-care test based in RT-LAMP | SARS-CoV-2 virus | 34 | 18 | 91.0 % | 100 % | NI | < 20 | Rodriguez et al. (2021) |
| Point-of-care lateral flow immunoassay | Serologic IgG and IgM | 397 | 128 | 88.66 % | 90.63% | NI | 15 | Li et al. (2020) |
| ME biosensor | Serologic IgG and IgM | 10 | 10 | 100 % | 100 % | 100 % | 1 | This work |

Since the COVID-19 pandemic, the need to develop new diagnostic platforms like biosensors for infectious diseases has become even more evident to provide more efficient diagnoses, helping, in particular, to reduce the spread of viral diseases (Shen et al., 2024; Kabay et al., 2022). The COVID-19 pandemic represented an unprecedented challenge for science, culminating in the publication of thousands of research articles in a few years (Riccaboni et al., 2022), which makes it a key and well-characterized disease to apply as a model for the development of a new biosensing platform. In this context, magnetoelastic-based sensors emerge as an excellent base platform for biosensing due to their versatility, low cost and robustness, with great potential for identifying various targets and biomarkers. Given this scenario, our work proposes a new functionalized magnetoelastic biosensor with SARS-CoV-2 N-nucleocapsid phosphoprotein antigen for COVID-19 detection associated with antibodies in human plasma.

## 2. Materials and methods

### 2.1. Obtaining, heterologous expression and purification of the recombinant protein N

The optimized recombinant N-nucleocapsid phosphoprotein (NCBI sequence gene ID: Gene ID: 43740575) with the addition of a N-terminal His-tag and gene sequence with codon optimization for expression in *Escherichia coli* was commercially acquired and inserted into the pET28a expression vector (Biomatik, Canada). Competent *Escherichia coli* DH5α cells were transformed by heat shock with the recombinant plasmid, the transformants were selected, the plasmids were extracted and verified by colony PCR, restriction enzyme digestion and sequencing.

 The confirmed recombinant plasmids were transformed into the *E. coli* Artics express strain for optimized recombinant N-nucleocapsid phosphoprotein expression. The *E. coli* Artics express containing the recombinant pET28a (+) were grown in Luria Bertani (LB) broth supplemented with kanamycin, and the protein expression was then induced by adding 1 mM isopropyl-β-d-thiogalactopyranoside (IPTG). The bacterial culture was centrifuged and the pellets were lysed by sonication under denaturing conditions using the binding buffer composed of 20 mM imidazole, 20 mM $NaH_2PO_4$, and 0.5 M NaCl. After new centrifugation, the insoluble fraction was then collected, solubilized in solubilization buffer (binding buffer added 8 M urea), and applied to Ni-NTA columns (Qiagen, Hilden, Germany) in ÄKTA-Start™ System (GE Healthcare Life Sciences) for purification of His-tagged protein using binding and elution buffer (20 mM $NaH_2PO_4$, 0.5 mM NaCl and 500 mM imidazole).

The purified recombinant N-nucleocapsid phosphoprotein was analyzed by SDS-PAGE electrophoresis and Western blotting to evaluate their identity and purity. For Sodium dodecyl-sulfate polyacrylamide gel electrophoresis (SDS-PAGE), the N-nucleocapsid phosphoprotein fractions were mixed with the loading buffer (0.5 M Tris–HCl pH 6.8, 10% SDS, glycerol, bromophenol blue, DTT), heated at 95 °C for 10 minutes and subjected to 12.5% polyacrylamide gel electrophoresis (180 minutes at 95 V). The bands were stained using Coomassie Brilliant Blue dye (BioRad, UK). For WB analysis, the purified recombinant N-nucleocapsid phosphoprotein was subjected to new 12.5% polyacrylamide gel electrophoresis and the bands were transferred to a nitrocellulose membrane (BioRad, UK) at 400 mA for 1 hour. The membranes were blocked with a blocking buffer (1% Bovine Serum Albumin [BSA] in phosphate saline buffer with 0.05% Tween 20 [PBST]) for 12 hours at 4 °C. After three rounds of PBST washing, the membrane was incubated with conjugated anti-poly His antibody for 90 minutes at 4 °C, under shaking. Finally, the bands were visualized by adding the 3, 3'-diaminobenzidine tetrahydrochloride (DAB) solution (for 10 mL of DAB solution: 10 mg of DAB, 1 mL of 1 M Tris-HCl pH 7.6, 1 mL of 0.3% $NiCl_2$, 10 µL of $H_2O_2$ and $H_2O$).

### 2.2. Plasma samples and ELISA

The plasma samples used in this study belong to the plasma library of the Biotechnology and Molecular Biology Laboratory at the Federal University of Viçosa. Twenty plasma samples were used in this work, in which ten plasma samples were collected from patients infected with the SARS-CoV-2 virus confirmed by the real-time reverse transcription PCR (RT-PCR) assay using nasopharyngeal swabs, and ten other plasma samples from negative patients. A total of five plasma samples of patients infected with Influenza A (H1N1), Influenza A (H3N2), Influenza B, and Parainfluenza type 4 obtained from Fiocruz - Rene Rachou Institute were also used to evaluate the selectivity of antigen. The heat-inactivated plasma samples were obtained in October 2020 during the COVID-19 pandemic, from patients admitted to the Materdei and Risoleta Tolentino Neve Hospitals (Belo Horizonte, Minas Gerais) and are registered in the Human Research Ethical Commission in Rene Rachou Institute (CAAE 30399620.0.0000.5091 - registration 4.210.316).

To establish a comparison of the biosensor serodiagnostic capacity under development, plasma samples were firstly tested by the gold standard method enzyme-linked immunosorbent assay (ELISA) using the recombinant N-nucleocapsid phosphoprotein as antigen. For this, the wells of a 96-well plate were sensitized with 2 µg of recombinant N-nucleocapsid phosphoprotein diluted in coating buffer (15 mM $Na_2CO_3$, 34 mM $NaHCO_3$ pH 9.3) and incubated overnight at 4 °C. The coating buffer was washed four rounds with PBST and the wells were blocked using the blocking buffer (2% BSA in PBST). After incubation for 1 hour at 37 °C, the blocking buffer was removed and plasma diluted 1:400 in PBST was added, corresponding to the highest plasma dilution that was capable of clearly separating positive and negative patients for COVID-19 (Fig. S1). After another incubation for 1 hour at 37 °C, the wells were washed again with PBST, and anti-human IgG secondary antibodies in 1:20,000 titer were added to each well and incubated for 1 hour at 37 °C. The 3,3',5,5'-Tetramethylbenzidine solution (TMB) was added to the wells and kept for 5 minutes at room temperature in the dark and the reaction was then stopped by adding 0.5 M $H_2SO_4$. Finally, a microplate reader measured the optical density values at 450 nm (BioTek, USA).

### 2.3. Preparation of the ME sensor

Magnetoelastic strip-shaped resonator platforms, of 1 mm × 5 mm × 28 µm, were fabricated from METGLAS® 2826 MB3 alloy ($Fe_{40}Ni_{38}Mo_4B_{18}$). The ME ribbons were diced into rectangular shaped platforms using scissors and millimeter paper, which allowed precise cuts. The sensors were ultrasonically cleaned in acetone (5 min) followed by ethanol (5 min) and dried with nitrogen gas. Subsequently, on one side of the sensor, they received a 50 nm thick chromium (Cr) layer applied by magnetron sputtering technique. Later, they were coated with a 100 nm thick gold (Au (111)) layer using thermal evaporation. In these stages, was used an ATC Polaris 5 Sputter system and an Edwards Auto306 Turbo Evaporator equipped with an Edwards FTM6 quartz crystal microbalance to determine film thickness. The Cr layer promotes higher adhesion between the Au film and the sensor. The Au film provides a highly biocompatible surface, promoting increased adsorption of functionalizing biomolecules (Calzolari et al., 2007). Afterward, the ME ribbons were annealed in a vacuum (~$10^{-3}$ Torr) oven at 200 °C for 2 hours to relieve residual internal stress and promote the complete adhesion of the Au layer.

### 2.4. Antigen immobilization

The surface of the gold-impregnated biosensor was biofunctionalized with 2 µg of recombinant N-nucleocapsid phosphoprotein Fig. 1a. The biosensor was incubated in solution for 1 hour at room temperature. Then, the biosensor was washed with distilled water, dried on absorbent paper, and incubated in a blocking buffer (1% BSA in PBST) for 1 hour at room temperature. Once biofunctionalized and blocked, the biosensors were exposed to plasma obtained from patients confirmed or not for COVID-19 by RT-PCR, diluted in PBST at the same concentration used in the ELISA assay. The measurement of the magnetoelastic resonance frequency was carried out in solution, at room temperature, immediately after the exposition of the functionalized sensor (0 min), whose value we denote by $f_0$, and after incubation for 60 minutes, denoted by $f$. The frequency shift is given by $f_0 - f$.

### 2.5. Biochemical and surface functionalization validation

To confirm the immobilization of the recombinant antigen on the surface of the biosensor, scanning electron microscopy (SEM), energy-dispersive X-ray spectroscopy (EDX), atomic force microscopy (AFM), and micro-Raman spectroscopy analyses were used to evaluate the functionalization steps. SEM analysis was performed at an accelerating voltage of 15 kV and EDX spectra for elemental analysis (Au) of the biosensor surface before and after protein N immobilization were measured, using an accelerating voltage of 7 kV, both measures using the JEOL-JSM-6010LA microscope (JEOL Corporation, Tokyo, Japan). AFM analysis was performed to examine the immobilization effect of the antigen on the biosensor using the NT-MDT Integra Prima scanning probe microscope (NT-MDT, Zelenograd, Russia) at 23 ºC and 30% relative humidity operated in the tapping mode, obtaining images of 2 µm × 2 µm of area at a resolution of 256 × 256 pixels. The micro-Raman spectroscopy measurements were performed in an InVia Renishaw spectrometer (Renishaw, Watton-Under-Edge, United Kingdom) using a 785 nm excitation laser line focused by a 50X objective lens. All measurements were collected with 15 s acquisition time and 4 spectral accumulations. The spectra were deconvoluted using Fityk software (Wojdyr, 2010).

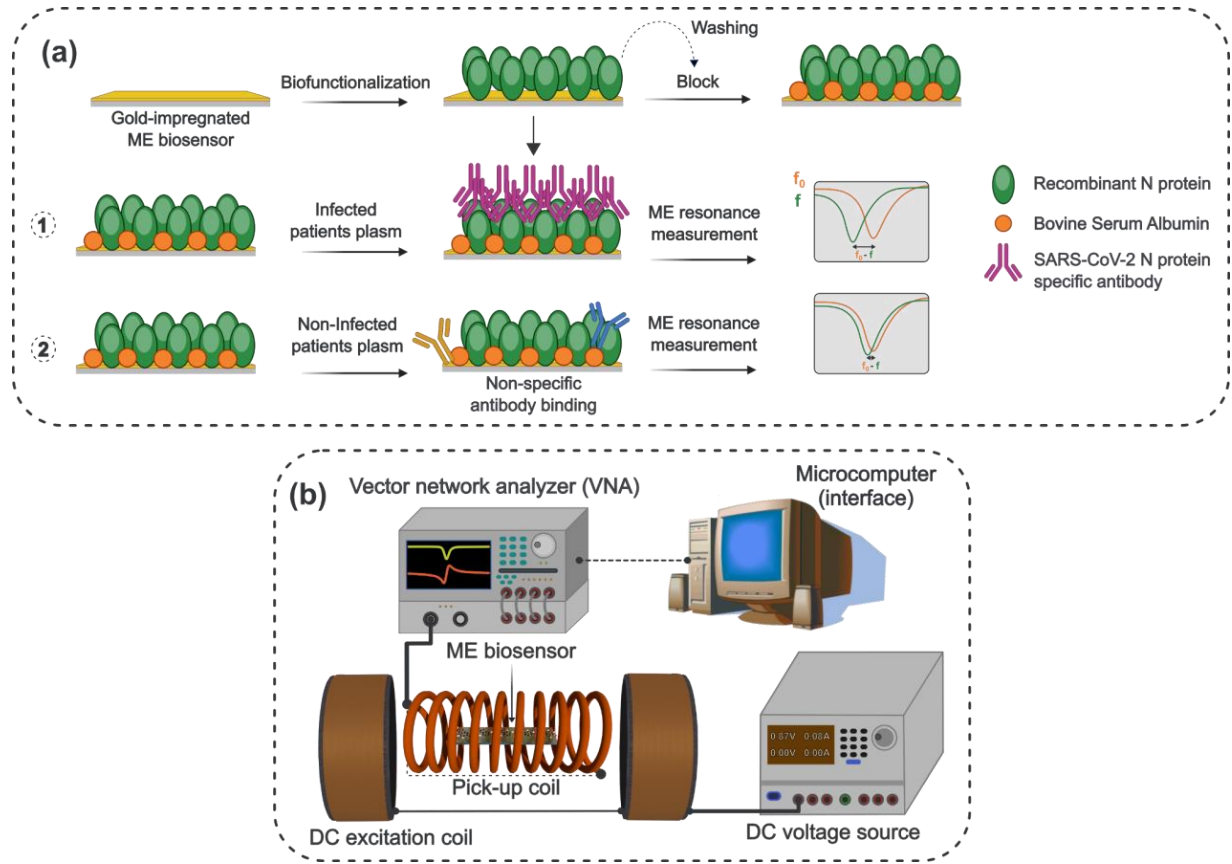

**Figure 1** - (a) scheme illustrating the functionalization steps of the ME biosensor. (b) Schematic of the experimental setup used in the process of measuring the magnetoelastic resonance frequency.

### 2.6. Principle of detection and resonance frequency measurements of ME biosensors

Due to a strong coupling between mechanical and magnetic energies, ME materials can change their dimensions in the presence of a magnetic field, whose deflections can be studied by evaluating the modulus of elasticity $E$ of the ME material. The modulus of elasticity is a function of the applied magnetic field, $E = E(H)$, by equation (1) (Bergmair et al., 2014; Cullity et al., 2011; Shen et al., 2010; Spano et al., 1982):

$$E(H) = \left( \frac{1}{E_H} + \frac{9\lambda_s{}^2 H^2}{\mu_0 M_s H_{A\sigma}^3} \right)^{-1}, \qquad (1)$$

where $E_H$ is the modulus of elasticity without the effect of the magnetic field ($H = 0$), $\mu_0$ is the vacuum magnetic permeability, $H$ is the applied magnetic field, $\lambda_s$ is the magnetostriction constant, $M_s$ is the saturation magnetization, and $H_{A\sigma}$ is the magnetoelastic anisotropy field, which depends on the stress $\sigma$ applied to the material.

When such ME materials are exposed to AC (modulation) and DC (induction) magnetic fields, a magnetoelastic wave propagates through the material, whose magnetoelastic resonance frequency $f_n$ can be located by equation (2) (Saiz et al. 2022):

$$f_n = \frac{n}{2L} \sqrt{\frac{E_H}{\rho(1-v^2)}}, \qquad (2)$$

where $n$ is the resonance mode (evaluated at $n = 1$), $L$ is the sensor length of density $\rho$ and Poisson's ratio $v$.

An increase in mass on the sensor surface causes a translation $\Delta f$ of the resonance frequency to lower frequencies. The change in resonance frequency is influenced by the thickness of the accumulated mass, given by (Sauerbrey 1959):

$$\Delta f = -\frac{\rho_m h}{2\rho d} f_0, \qquad (3)$$

where $\rho_m$ and $h$ are, respectively, the density and thickness of the deposited mass layer, $d$ and $f_0$ are the thickness and the fundamental resonance frequency of the ME sensor without adding mass, respectively. Assuming that the increase in mass $\Delta m$ occurs over the entire surface area of the ME sensor, equation (3) takes the form:

$$\Delta f = -f_0 \frac{\Delta m}{2m_0}, \qquad (4)$$

where $m_0$ and $f_0$ are the mass and resonance frequency of the bare magnetoelastic platform. Higher resonance frequencies, which can be achieved with sensors of smaller length $L$, or sensors of smaller mass $m_0$ result in higher sensitivity value. The mass sensitivity is defined in equation (5) (Saiz et al. 2022).

$$s_m = -\frac{\Delta f}{\Delta m} = \frac{f_0}{2m_0}. \qquad (5)$$

Resonance frequency measurements of the ME biosensors were performed using a vector network analyzer (R&S®ZNLE18, Rohde & Schwarz, Munich, Germany), operated in $S_{11}$ mode connected to a double-layer excitation/pickup solenoid coil (200 turns, 4 mm inner diameter and 22 mm long, with 0.18 mm thick copper wire and 4.5 $\Omega$ resistance), which generates the AC excitation signal with 5 Hz step, to perform a frequency sweep and monitor the reflected signal (see Fig.1b). The DC magnetic induction field was applied through a second solenoid coil (600 turns, 23 mm internal diameter and 50 mm long, with 0.40 mm thick copper wire and 10.7 $\Omega$ resistance) concentric to the excitation/capture coil, using a Keysight U8031A DC source. The configuration of the excitation/pickup coils and the DC coil was chosen carefully to maintain an appropriate number of windings, thus minimizing the increase in coil resistance. The attention to this detail is essential for mitigating the rise in thermal noise, which can adversely affect the signal-to-noise ratio. Each sensor was placed inside an Eppendorf microtube with 0.5 mL of the analyzed solution, which was inserted vertically inside the excitation/capture coil to measure the magnetoelastic resonance frequency. All measurements were carried out at room temperature (25 °C). The resonant frequency of the biosensor is determined by measuring parameter $S_{11}$, which was monitored and recorded every 10 minutes.

### 2.7 Statistical Analysis

Statistical analyses were performed using GraphPad Prism (version 8.3). The cutoff value, the separation limit between positive and negative results for the patients' plasma, was calculated as the sum of the means of the negative values plus twice the standard deviation of the mean (Greiner et al., 1995). The performance of each test was evaluated according to sensitivity, specificity, positive predictive value (PPV), negative predictive value (NPV), and accuracy using the MedCalc online platform (https://www.medcalc.org/calc/diagnostic_test.php).

## 3. Results and discussion

### 3.1. Assessment of the surface functionalization of gold.

In Fig. 2a is the SDS-PAGE electrophoresis to confirm the antigen purity and the sequence identity by Western blotting using commercial antibodies anti-His Tag and antibodies anti- N-nucleocapsid phosphoprotein. The EDX spectra for elemental analysis of the biosensor surface before and after immobilization of the N-nucleocapsid phosphoprotein are presented in Fig. 2b. Protein N is a macromolecule made up of chains of amino acids, containing large amounts of carbon (C), nitrogen (N), oxygen (O) and hydrogen (H), which were evidenced in EDX measurements, except for H, due to the limitation of the microscope in detecting it. Such elements were also reported by Zhang et al. (2023). A remaining amount of NaCl is also found, resulting from the N protein purification process. Fig. 2c shows the Raman spectrum of the solution before functionalization to characterize the antigen signal. Based on the previous work of Sanchez, we assigned the Raman peaks located between ~500 cm$^{-1}$ and 1700 cm$^{-1}$ as belonging to the N-nucleocapsid phosphoprotein (Sanchez et al., 2021). The same vibrational modes are observed for the functionalized surface Fig. 2d, confirming the attachment of the N-nucleocapsid phosphoprotein to the Au film. Given the analysis presented in Fig. 2b-d, it can be concluded that the N-nucleocapsid phosphoprotein is successfully immobilized on the surface of the biosensor.

The potential of antibodies from COVID-19 human samples to interact with functionalized sensors was also analyzed with AFM and SEM, as shown in Fig. 3. After the functionalization process, the sensors were dried for 1 hour in a stereo environment and blown with a nitrogen gas jet at the end, and then taken for analysis. Characterization of the biosensor surface morphology was performed to examine the effect of immobilization of protein together with the potential reactivity with human positive and negative samples for anti-SARS-CoV-2 antibodies using functionalized devices. Fig. 3a-c present the AFM images of the surface for ME biosensors. Figure 3a shows the Au surface coated with the N-nucleocapsid phosphoprotein antigen + blocking BSA protein, where a maximum height of 8.5 nm is observed. Figure 3b shows the surface of the platform after modification with negative serum, showing a maximum height of 31 nm, and Fig. 3c refers to the positive serum, with a maximum height of 69 nm. Compared with the surface height of the Au coated with the N-nucleocapsid phosphoprotein antigen + blocking BSA protein, an increase of approximately 23 nm is observed for the negative serum and 61 nm for the positive human plasma. The increase in height indicates a change in the surface morphology of the ME biosensor and it is attributed to the physical adsorption of serum biomolecules selectively chosen by the N-nucleocapsid phosphoprotein and fixed on its surface (Wang et al., 2020; Sang et al., 2018; Horikawa et al., 2011).

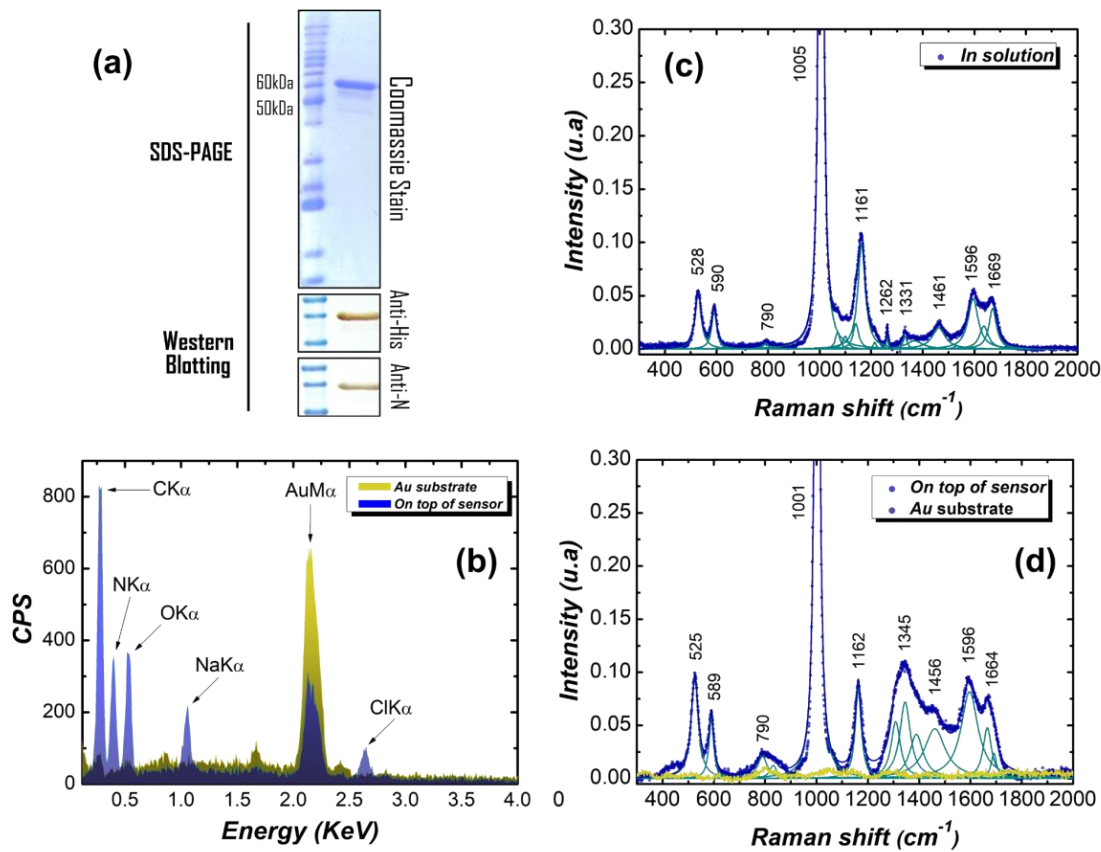

**Figure 2** - (a) SDS-PAGE electrophoresis and Western blotting using commercial antibodies anti-His tag and Anti-N-nucleocapsid phosphoprotein. (b) EDX spectrum of the recombinant N-nucleocapsid phosphoprotein on the sensor, after the functionalization process together with the signal from the Au substrate. (c) Raman spectrum of recombinant N-nucleocapsid phosphoprotein in solution and (d) Raman spectrum of the recombinant N-nucleocapsid phosphoprotein on the sensor, after the functionalization process together with the Raman signal from the Au substrate.

Figure 3d-f shows SEM images for ME biosensors of the Au surface coated with the N-nucleocapsid phosphoprotein + blocking BSA protein (3d), after interaction with the negative (3e) and positive (3f) human samples. A characteristic formation pattern of dendritic shape Fig. 3f in sensors exposed to positive plasma is observed. This layer indicates the deposit of COVID-19 antibodies selected by the N-nucleocapsid phosphoprotein. The images in Fig. 3a-b helped in the comparison process, but it was not possible to observe marked changes on the sensor surface, as expected. Overall, the AFM and SEM images demonstrate that the surface functionalization of the ME biosensor with specific interaction with positive human samples was successful.

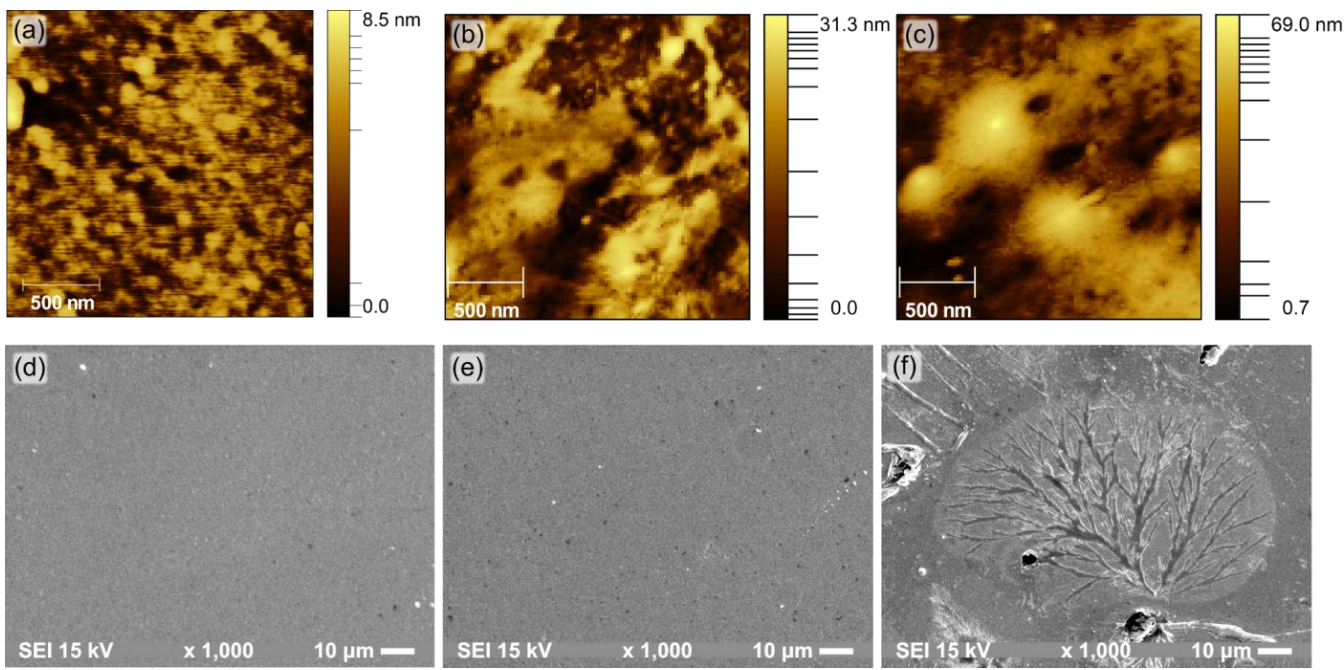

**Figure 3** - AFM images of the surface of the functionalized ME biosensor, captured from the sensor with the blocking protein (a), after interaction with the negative (b) and positive (c) serum. Similarly, the SEM images in the blocking step (d), after interaction with the negative (e) and positive (f) serum.

The results obtained in measurements of the shift in magnetoelastic resonance frequency Fig. 4 corroborate those obtained in morphological characterization using AFM and SEM, since the frequency shift also showed a notable increase in serum negative patients Fig. 4a compared to positive human samples Fig. 4b. This result is justified since serum positive human samples have specific antibodies against N-nucleocapsid phosphoprotein antigen, which were selectively deposited on the sensors, resulting in a more pronounced shift in the resonance frequency. The deviations in the resonance frequencies shown in Fig. 4c for plasma-negative patients indicate a background noise in the interaction between the N-nucleocapsid phosphoprotein and the antibodies of the negative serum, which differ significantly from the signal of plasma-positive samples.

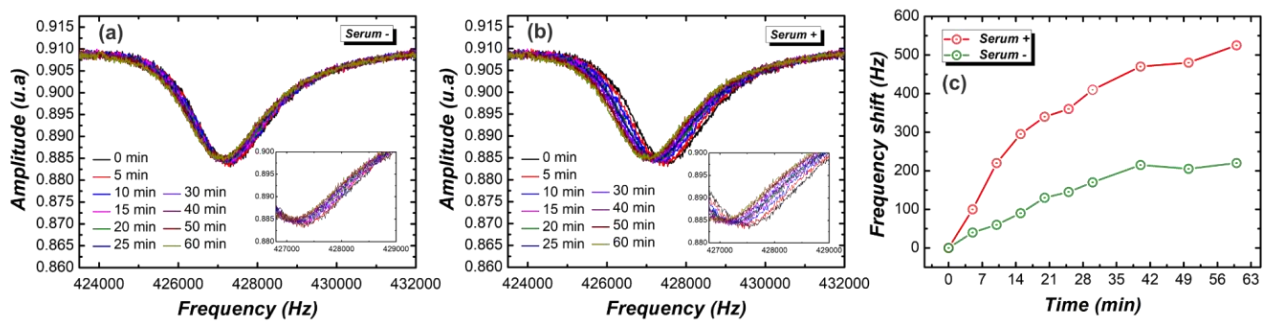

**Figure 4** - Magnetoelastic resonance spectra of the sensor exposed to the (a) negative and the (b) positive serum, during the functionalization time. The changes in the resonance frequency of the sensor exposed to positive and negative serum can be seen in image (c).

### 3.2. ME biosensors for anti-Sars-Cov-2 antibody detection

The shifts in resonance frequencies for ten sensors exposed to negative serum and ten sensors exposed to positive serum are shown in Fig 5a. The frequency sweep was carried out with a step of 5 Hz. A significant difference is noted between the frequency shift obtained from sensors exposed to negative serum in relation to positive samples (talk about cut-off), with an average difference between them of 427.1 Hz. Therefore, it is evident that the proposed sensor can be used to distinguish negative human plasma from positive samples. All samples were also subjected to gold standard ELISA test Fig. 5b, and the presence of specific antibodies was certified, providing robustness to the data obtained by the proposed magnetoelastic sensors. The sample titer (1:400) used in both ELISA and ME sensors was selected due to the optimized response detected against the recombinant antigen in the comparison between infected and control samples (see Fig. S1). Moreover, the recombinant antigen is specific for anti-Sars-Cov-2 antibody detection as observed in the selectivity test realized using samples from patients with different respiratory infectious diseases (see Fig. S2). The comparison of diagnostic performance Fig. 5c-d showed that the new ME biosensor has the same diagnostic power as ELISA validated by the same values obtained for all statistical parameters, including sensitivity, specificity, positive and negative predictive values and accuracy. Despite this, this new biosensor has advantages when compared to ELISA, including lower cost due to no need for a microplate, chemically labeled secondary antibodies and colorimetric reagent. Moreover, the use of the ME biosensor is faster to execute as it has a protocol with fewer steps compared to the ELISA method. Another advantage is the digitalization of signal obtained in the ME biosensor approach, which is more suitable for automating diagnostic tests and analyzing of large number of samples, as well as facilitating conversion to methods that require quantification of the analyte to be measured, since it does not require the production of a standard curve, generally necessary in colorimetric and fluorescent tests such as ELISA. This certification strengthens the reliability of the proposed sensor, validating its effectiveness in differentiating between negative and positive plasma samples. This certification provides reliability to the data obtained by magnetoelastic sensors, reinforcing their usefulness and filling a significant gap in existing SARS-CoV-2 and COVID-19 serodiagnosis approaches and the possibility to apply the same approach for other infection disease, including the possibility of the development of new devices point-of-care for rapid diagnosis.

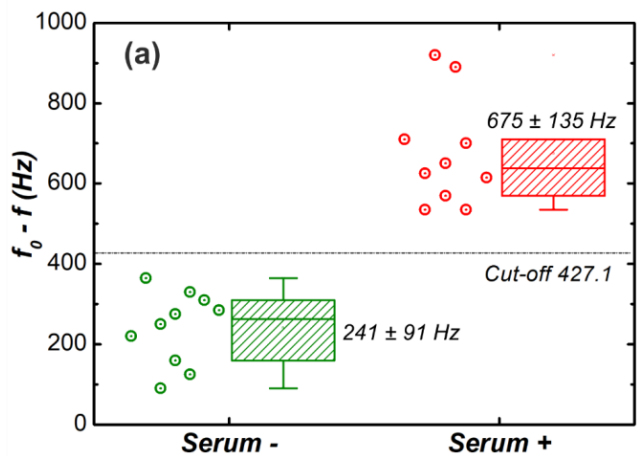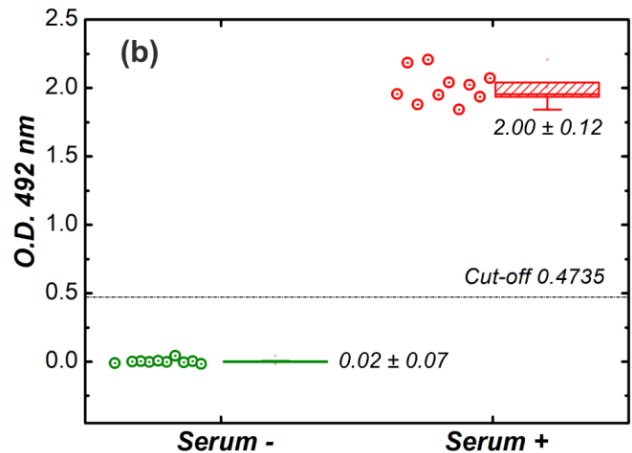

**(c)**

| Statistic | Value | 95% CI |
|---|---|---|
| Sensitivity | 100.00% | 65.00% to 100.00% |
| Specificity | 100.00% | 65.15% to 100.00% |
| Negative Lokelihood Ratio | 0.00 | |
| Disease prevalence (*) | 50.00% | 27.20% to 72.80% |
| Positive Predictive Value (*) | 100.00% | 69.15% to 100.00% |
| Negative Predictive Value (*) | 100.00% | 69.15% to 100.00% |
| Accuracy (*) | 100.00% | 83.16% to 100.00% |

(*) These values are dependent on disease prevalence

**(d)**

| Statistic | Value | 95% CI |
|---|---|---|
| Sensitivity | 100.00% | 65.00% to 100.00% |
| Specificity | 100.00% | 65.15% to 100.00% |
| Negative Lokelihood Ratio | 0.00 | |
| Disease prevalence (*) | 50.00% | 27.20% to 72.80% |
| Positive Predictive Value (*) | 100.00% | 69.15% to 100.00% |
| Negative Predictive Value (*) | 100.00% | 69.15% to 100.00% |
| Accuracy (*) | 100.00% | 83.16% to 100.00% |

(*) These values are dependent on disease prevalence

**Figure 5** - (a) Resonance frequency shifts measured with the average value and standard deviation (SD) for the twenty sensors functionalized with the N-nucleocapsid phosphoprotein after exposure to positive (red) and negative plasm (green). (b) Measured the optical density values after ELISA validated positive (red) and negative plasm (green), with the average value and standard deviation (SD). Sensitivity, Specificity, positive predictive value, negative predictive value and accuracy of measurements performed with the (c) biosensor and (d) ELISA, respectively.

## 4. Conclusion

In conclusion, the study presents a new efficient magnetoelastic biosensor, functionalized with the N antige N-nucleocapsid phosphoprotein of the SARS-CoV-2 virus for the detection of antibodies in human plasma, aiming at the serological diagnosis of COVID-19. Comprehensive validation, including techniques such as SDS-PAGE, Western blotting, atomic force microscopy (AFM), scanning electron microscopy (SEM), energy-dispersive X-ray (EDX) microanalysis, and micro-Raman spectroscopy, confirm the selectivity and effective surface functionalization of the biosensor. The results indicate the robustness of the ME biosensor in reliably differentiating between positive and negative samples, with performance equivalent to the enzyme-linked immunosorbent assays (ELISA) method. The ME biosensor has the potential for automation in large-scale diagnostic studies, as a significant promise to fill the gap in SARS-CoV-2 detection approaches, standing out as an efficient, low-cost and promising tool for the serological diagnosis of COVID-19.

**Acknowledgements**

This research is supported by Conselho Nacional de Desenvolvimento Científico e Tecnológico (CNPq), Coordenação de Aperfeiçoamento de Pessoal de Nível Superior (CAPES), Financiadora de Estudos e Projetos (FINEP), Fundação de Amparo à Pesquisa do Estado de Minas Gerais (FAPEMIG) - Rede de Pesquisa em Materiais 2D and Rede de Nanomagnetismo, and INCT of Spintronics and Advanced Magnetic Nanostructures (INCT-SpinNanoMag), No. CNPq 406836/2022-1.

# SUPPLEMENTARY MATERIAL

The supplementary material presents additional results on the enzyme-linked immunosorbent assay for COVID-19 and antigen selectivity. With these results, experimental conditions can be established that maximize sensitivity and specificity to detect antibodies related to SARS-CoV-2, minimizing cross-reactions with other respiratory diseases.

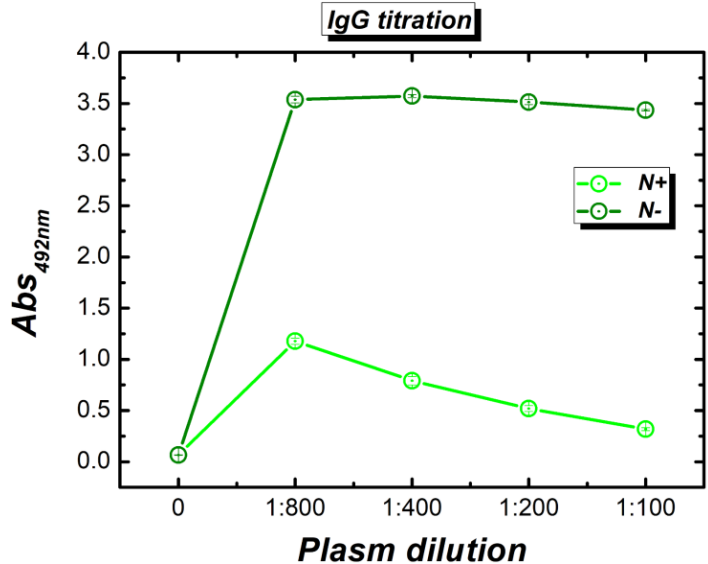

Figure S1 - Titration assay with plasma from positive (N+) and negative (N-) patients for COVID-19 by ELISA immunoenzymatic assay. The highest serum dilution that was shown to be capable of clearly separating positive and negative patients for COVID-19 was chosen for all subsequent experiments (1:400).

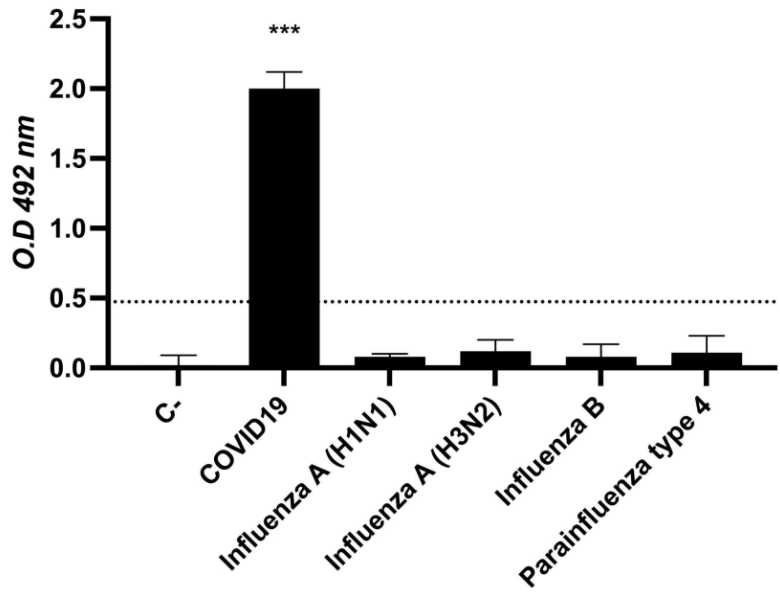

Figure S2 - ELISA immunoenzymatic assay with plasma from patients with different respiratory infectious diseases to evaluate the selectivity of recombinant N-nucleocapsid phosphoprotein used as antigen. C-, patients with no history of respiratory disease in the last 6 months. The dotted line represents the cut off value (O.D. 492 nm = 0.4735).